\def\epsfigload{1}

\def\cerntp{1}

\def\pltp{0}

\ifnum\epsfigload=1
\documentstyle[epsfig,ifthen]{dgelsart}
\else
\documentstyle[ifthen]{dgelsart}
\newcommand{\psdraft}{}
\newcommand{\epsfig}[1]{}
\fi

\newcommand{\ifdraft}{nodraft}

\newcommand{\ifpsdraft}{nodraft}

\newcommand{\vdate}{June 1996}
\newcommand{\cernnr}{96--155}
\newcommand{\beq}{\begin{equation}}
\newcommand{\eeq}{\end{equation}}
\newcommand{\beqn}{\begin{eqnarray}}
\newcommand{\eeqn}{\end{eqnarray}}

\newcommand{\GeV}{\mbox{GeV}}

\newcommand{\dd}{\mbox{d}}

\newcommand{\dgfootnote}[1]{$\!$\footnote{#1}}

\newcommand{\fullline}{
\unitlength0.4mm
\begin{picture}(13,4)
\linethickness{0.3mm}
\put(-1,2.0){\line(1,0){15}}
\thinlines
\end{picture}
}

\newcommand{\dashline}{
\unitlength0.4mm
\begin{picture}(20,4)
\linethickness{0.3mm}
\put(-1,2.0){\line(1,0){4}}
\put(8,2.0){\line(1,0){4}}
\put(17,2.0){\line(1,0){4}}
\thinlines
\end{picture}
}

\newcommand{\dotline}{
\unitlength0.4mm
\begin{picture}(9,4)
\linethickness{0.3mm}
\put(-1,2.0){\line(1,0){1}}
\put(4,2.0){\line(1,0){1}}
\put(9,2.0){\line(1,0){1}}
\thinlines
\end{picture}
}

\newcommand{\longdashline}{
\unitlength0.4mm
\begin{picture}(22,4)
\linethickness{0.3mm}
\put(-1,2.0){\line(1,0){10}}
\put(13,2.0){\line(1,0){10}}
\thinlines
\end{picture}
}

\newcommand{\dashdotline}{
\unitlength0.4mm
\begin{picture}(17,4)
\linethickness{0.3mm}
\put(-1,2.0){\line(1,0){5}}
\put(8,2.0){\line(1,0){1}}
\put(13,2.0){\line(1,0){5}}
\thinlines
\end{picture}
}

\newcommand{\dotdotline}{
\unitlength0.4mm
\begin{picture}(12,4)
\linethickness{0.3mm}
\put(-1,2.0){\line(1,0){1}}
\put(2,2.0){\line(1,0){1}}
\put(9,2.0){\line(1,0){1}}
\put(12,2.0){\line(1,0){1}}
\thinlines
\end{picture}
}

\newcommand{\dashdotdotline}{
\unitlength0.4mm
\begin{picture}(11,4)
\linethickness{0.3mm}
\put(-1,2.0){\line(1,0){5}}
\put(8,2.0){\line(1,0){1}}
\put(11,2.0){\line(1,0){1}}
\thinlines
\end{picture}
}

\newcommand{\dgpicture}[2]{
\begin{picture}(#1,#2)
\thicklines
\thinlines
}

\newcommand{\epsfigdg}[2]{\epsfig{figure=#1,#2}}

\newcommand{\chsign}[1]{%
{\ifthenelse{\equal{\ifdraft}{draft}}%
{
{\sf {\Large$\bullet$$\bullet$$\bullet$} #1 
                         {\Large$\bullet$$\bullet$$\bullet$} }%
}%
{}%
}}
\newcommand{\smallmark}[1]{
\marginpar{\fbox{\vspace{0.0cm}{\scriptsize #1}}}}

\newcommand{\labelm}[1]{%
\label{#1}%
\ifthenelse{\equal{\ifdraft}{draft}}%
{\smallmark{#1}}%
{}%
}

\newcommand{\labelmm}[1]{%
\label{#1}%
\ifthenelse{\equal{\ifdraft}{draft}}%
{\protect\fbox{\sf #1}}%
{}%
}

\newcommand{\beqm}[1]{%
\ifthenelse{\equal{\ifdraft}{draft}}%
{\smallmark{#1}}%
{}%
\beq \label{#1}}

\newcommand{\beqnm}[1]{%
\ifthenelse{\equal{\ifdraft}{draft}}%
{\smallmark{#1}}%
{}%
\beqn \label{#1}}

\ifthenelse{\equal{\ifpsdraft}{draft}}{\psdraft}{}

\catcode`\@=11

 \font\tenmsx=msam10 scaled \magstep1
 \font\sevenmsx=msam8
 \font\fivemsx=msam6
 \font\tenmsy=msbm10 scaled \magstep1
 \font\sevenmsy=msbm8
 \font\fivemsy=msbm6

\newfam\msxfam
\newfam\msyfam
\textfont\msxfam=\tenmsx  \scriptfont\msxfam=\sevenmsx
  \scriptscriptfont\msxfam=\fivemsx
\textfont\msyfam=\tenmsy  \scriptfont\msyfam=\sevenmsy
  \scriptscriptfont\msyfam=\fivemsy

\def\hexnumber@#1{\ifnum#1<10 \number#1\else
 \ifnum#1=10 A\else\ifnum#1=11 B\else\ifnum#1=12 C\else
 \ifnum#1=13 D\else\ifnum#1=14 E\else\ifnum#1=15 F\fi\fi\fi\fi\fi\fi\fi}

\def\msx@{\hexnumber@\msxfam}
\def\msy@{\hexnumber@\msyfam}
\mathchardef\boxdot="2\msx@00
\mathchardef\boxplus="2\msx@01
\mathchardef\boxtimes="2\msx@02
\mathchardef\square="0\msx@03
\mathchardef\blacksquare="0\msx@04
\mathchardef\centerdot="2\msx@05
\mathchardef\lozenge="0\msx@06
\mathchardef\blacklozenge="0\msx@07
\mathchardef\circlearrowright="3\msx@08
\mathchardef\circlearrowleft="3\msx@09
\mathchardef\rightleftharpoons="3\msx@0A
\mathchardef\leftrightharpoons="3\msx@0B
\mathchardef\boxminus="2\msx@0C
\mathchardef\Vdash="3\msx@0D
\mathchardef\Vvdash="3\msx@0E
\mathchardef\vDash="3\msx@0F
\mathchardef\twoheadrightarrow="3\msx@10
\mathchardef\twoheadleftarrow="3\msx@11
\mathchardef\leftleftarrows="3\msx@12
\mathchardef\rightrightarrows="3\msx@13
\mathchardef\upuparrows="3\msx@14
\mathchardef\downdownarrows="3\msx@15
\mathchardef\upharpoonright="3\msx@16

\mathchardef\downharpoonright="3\msx@17
\mathchardef\upharpoonleft="3\msx@18
\mathchardef\downharpoonleft="3\msx@19
\mathchardef\rightarrowtail="3\msx@1A
\mathchardef\leftarrowtail="3\msx@1B
\mathchardef\leftrightarrows="3\msx@1C
\mathchardef\rightleftarrows="3\msx@1D
\mathchardef\Lsh="3\msx@1E
\mathchardef\Rsh="3\msx@1F
\mathchardef\rightsquigarrow="3\msx@20
\mathchardef\leftrightsquigarrow="3\msx@21
\mathchardef\looparrowleft="3\msx@22
\mathchardef\looparrowright="3\msx@23
\mathchardef\circeq="3\msx@24
\mathchardef\succsim="3\msx@25
\mathchardef\gtrsim="3\msx@26
\mathchardef\gtrapprox="3\msx@27
\mathchardef\multimap="3\msx@28
\mathchardef\therefore="3\msx@29
\mathchardef\because="3\msx@2A
\mathchardef\doteqdot="3\msx@2B

\mathchardef\triangleq="3\msx@2C
\mathchardef\precsim="3\msx@2D
\mathchardef\lesssim="3\msx@2E
\mathchardef\lessapprox="3\msx@2F
\mathchardef\eqslantless="3\msx@30
\mathchardef\eqslantgtr="3\msx@31
\mathchardef\curlyeqprec="3\msx@32
\mathchardef\curlyeqsucc="3\msx@33
\mathchardef\preccurlyeq="3\msx@34
\mathchardef\leqq="3\msx@35
\mathchardef\leqslant="3\msx@36
\mathchardef\lessgtr="3\msx@37
\mathchardef\backprime="0\msx@38
\mathchardef\risingdotseq="3\msx@3A
\mathchardef\fallingdotseq="3\msx@3B
\mathchardef\succcurlyeq="3\msx@3C
\mathchardef\geqq="3\msx@3D
\mathchardef\geqslant="3\msx@3E
\mathchardef\gtrless="3\msx@3F
\mathchardef\sqsubset="3\msx@40
\mathchardef\sqsupset="3\msx@41
\mathchardef\vartriangleright="3\msx@42
\mathchardef\vartriangleleft="3\msx@43
\mathchardef\trianglerighteq="3\msx@44
\mathchardef\trianglelefteq="3\msx@45
\mathchardef\bigstar="0\msx@46
\mathchardef\between="3\msx@47
\mathchardef\blacktriangledown="0\msx@48
\mathchardef\blacktriangleright="3\msx@49
\mathchardef\blacktriangleleft="3\msx@4A
\mathchardef\vartriangle="3\msx@4D
\mathchardef\blacktriangle="0\msx@4E
\mathchardef\triangledown="0\msx@4F
\mathchardef\eqcirc="3\msx@50
\mathchardef\lesseqgtr="3\msx@51
\mathchardef\gtreqless="3\msx@52
\mathchardef\lesseqqgtr="3\msx@53
\mathchardef\gtreqqless="3\msx@54
\mathchardef\Rrightarrow="3\msx@56
\mathchardef\Lleftarrow="3\msx@57
\mathchardef\veebar="2\msx@59
\mathchardef\barwedge="2\msx@5A
\mathchardef\doublebarwedge="2\msx@5B
\mathchardef\angle="0\msx@5C
\mathchardef\measuredangle="0\msx@5D
\mathchardef\sphericalangle="0\msx@5E
\mathchardef\varpropto="3\msx@5F
\mathchardef\smallsmile="3\msx@60
\mathchardef\smallfrown="3\msx@61
\mathchardef\Subset="3\msx@62
\mathchardef\Supset="3\msx@63
\mathchardef\Cup="2\msx@64

\mathchardef\Cap="2\msx@65

\mathchardef\curlywedge="2\msx@66
\mathchardef\curlyvee="2\msx@67
\mathchardef\leftthreetimes="2\msx@68
\mathchardef\rightthreetimes="2\msx@69
\mathchardef\subseteqq="3\msx@6A
\mathchardef\supseteqq="3\msx@6B
\mathchardef\bumpeq="3\msx@6C
\mathchardef\Bumpeq="3\msx@6D
\mathchardef\lll="3\msx@6E

\mathchardef\ggg="3\msx@6F

\mathchardef\circledS="0\msx@73
\mathchardef\pitchfork="3\msx@74
\mathchardef\dotplus="2\msx@75
\mathchardef\backsim="3\msx@76
\mathchardef\backsimeq="3\msx@77
\mathchardef\complement="0\msx@7B
\mathchardef\intercal="2\msx@7C
\mathchardef\circledcirc="2\msx@7D
\mathchardef\circledast="2\msx@7E
\mathchardef\circleddash="2\msx@7F
\def\ulcorner{\delimiter"4\msx@70\msx@70 }
\def\urcorner{\delimiter"5\msx@71\msx@71 }
\def\llcorner{\delimiter"4\msx@78\msx@78 }
\def\lrcorner{\delimiter"5\msx@79\msx@79 }
\def\yen{\mathhexbox\msx@55 }
\def\checkmark{\mathhexbox\msx@58 }
\def\circledR{\mathhexbox\msx@72 }
\def\maltese{\mathhexbox\msx@7A }
\mathchardef\lvertneqq="3\msy@00
\mathchardef\gvertneqq="3\msy@01
\mathchardef\nleq="3\msy@02
\mathchardef\ngeq="3\msy@03
\mathchardef\nless="3\msy@04
\mathchardef\ngtr="3\msy@05
\mathchardef\nprec="3\msy@06
\mathchardef\nsucc="3\msy@07
\mathchardef\lneqq="3\msy@08
\mathchardef\gneqq="3\msy@09
\mathchardef\nleqslant="3\msy@0A
\mathchardef\ngeqslant="3\msy@0B
\mathchardef\lneq="3\msy@0C
\mathchardef\gneq="3\msy@0D
\mathchardef\npreceq="3\msy@0E
\mathchardef\nsucceq="3\msy@0F
\mathchardef\precnsim="3\msy@10
\mathchardef\succnsim="3\msy@11
\mathchardef\lnsim="3\msy@12
\mathchardef\gnsim="3\msy@13
\mathchardef\nleqq="3\msy@14
\mathchardef\ngeqq="3\msy@15
\mathchardef\precneqq="3\msy@16
\mathchardef\succneqq="3\msy@17
\mathchardef\precnapprox="3\msy@18
\mathchardef\succnapprox="3\msy@19
\mathchardef\lnapprox="3\msy@1A
\mathchardef\gnapprox="3\msy@1B
\mathchardef\nsim="3\msy@1C
\mathchardef\napprox="3\msy@1D
\mathchardef\varsubsetneq="3\msy@20
\mathchardef\varsupsetneq="3\msy@21
\mathchardef\nsubseteqq="3\msy@22
\mathchardef\nsupseteqq="3\msy@23
\mathchardef\subsetneqq="3\msy@24
\mathchardef\supsetneqq="3\msy@25
\mathchardef\varsubsetneqq="3\msy@26
\mathchardef\varsupsetneqq="3\msy@27
\mathchardef\subsetneq="3\msy@28
\mathchardef\supsetneq="3\msy@29
\mathchardef\nsubseteq="3\msy@2A
\mathchardef\nsupseteq="3\msy@2B
\mathchardef\nparallel="3\msy@2C
\mathchardef\nmid="3\msy@2D
\mathchardef\nshortmid="3\msy@2E
\mathchardef\nshortparallel="3\msy@2F
\mathchardef\nvdash="3\msy@30
\mathchardef\nVdash="3\msy@31
\mathchardef\nvDash="3\msy@32
\mathchardef\nVDash="3\msy@33
\mathchardef\ntrianglerighteq="3\msy@34
\mathchardef\ntrianglelefteq="3\msy@35
\mathchardef\ntriangleleft="3\msy@36
\mathchardef\ntriangleright="3\msy@37
\mathchardef\nleftarrow="3\msy@38
\mathchardef\nrightarrow="3\msy@39
\mathchardef\nLeftarrow="3\msy@3A
\mathchardef\nRightarrow="3\msy@3B
\mathchardef\nLeftrightarrow="3\msy@3C
\mathchardef\nleftrightarrow="3\msy@3D
\mathchardef\divideontimes="2\msy@3E
\mathchardef\varnothing="0\msy@3F
\mathchardef\nexists="0\msy@40
\mathchardef\mho="0\msy@66
\mathchardef\thorn="0\msy@67
\mathchardef\beth="0\msy@69
\mathchardef\gimel="0\msy@6A
\mathchardef\daleth="0\msy@6B
\mathchardef\lessdot="3\msy@6C
\mathchardef\gtrdot="3\msy@6D
\mathchardef\ltimes="2\msy@6E
\mathchardef\rtimes="2\msy@6F
\mathchardef\shortmid="3\msy@70
\mathchardef\shortparallel="3\msy@71
\mathchardef\smallsetminus="2\msy@72
\mathchardef\thicksim="3\msy@73
\mathchardef\thickapprox="3\msy@74
\mathchardef\approxeq="3\msy@75
\mathchardef\succapprox="3\msy@76
\mathchardef\precapprox="3\msy@77
\mathchardef\curvearrowleft="3\msy@78
\mathchardef\curvearrowright="3\msy@79
\mathchardef\digamma="0\msy@7A
\mathchardef\varkappa="0\msy@7B
\mathchardef\hslash="0\msy@7D
\mathchardef\hbar="0\msy@7E
\mathchardef\backepsilon="3\msy@7F
\def\Bbb{\ifmmode\let\next\Bbb@\else
 \def\next{\errmessage{Use \string\Bbb\space only in math mode}}\fi\next}
\def\Bbb@#1{{\Bbb@@{#1}}}
\def\Bbb@@#1{\fam\msyfam#1}

\catcode`\@=12
\font\teneusmf=eufm10 scaled 1200
\font\seveneusmf=eufm8
\font\fiveeusmf=eufm6
\newfam\eusmffam
\textfont\eusmffam=\teneusmf
\scriptfont\eusmffam=\seveneusmf
\scriptscriptfont\eusmffam=\fiveeusmf

\font\teneusm=eusm10 scaled 1200
\font\seveneusm=eusm8
\font\fiveeusm=eusm6
\newfam\eusmfam
\textfont\eusmfam=\teneusm
\scriptfont\eusmfam=\seveneusm
\scriptscriptfont\eusmfam=\fiveeusm

\font\teneusmc=cmsy10 scaled 1200
\font\seveneusmc=cmsy8
\font\fiveeusmc=cmsy6
\newfam\eusmcfam
\textfont\eusmcfam=\teneusmc
\scriptfont\eusmcfam=\seveneusmc
\scriptscriptfont\eusmcfam=\fiveeusmc

\newcommand{\titletextCERN}
{
Charged-Meson Production and\\ 
Scaling Violations of Fragmentation Functions\\
in Deeply Inelastic Scattering at HERA\\
}

\newcommand{\titletextPL}
{
Charged-Meson Production\\ 
and Scaling Violations of Fragmentation Functions\\
in Deeply Inelastic Scattering at HERA
}

\newcommand{\abstracttext}
{
We compare recent experimental results for one-particle-inclusive processes
in deeply inelastic scattering at HERA with theoretical predictions in 
next-to-leading-order QCD perturbation theory, and study the factorization 
scale dependence of cross sections and charged multiplicities. 
In the future, for the HERA machine running at design luminosity, scaling 
violations of fragmentation functions permit the measurement of the strong 
coupling constant. We estimate the size of the statistical error of $\alpha_s$
that can be achieved, and study the theoretical error due to the various 
parton density parametrizations and due to the factorization scale dependence.
}

\begin{document}


\ifnum\cerntp=1

\thispagestyle{empty}

\renewcommand{\thefootnote}{\fnsymbol{footnote}}
\setcounter{footnote}{0}


\begin{flushright}
{
\unitlength 1mm
\begin{picture}(10,10)
\put(0,0){CERN--TH/\cernnr}
\put(0,-5){hep-ph/9606470}
\end{picture}
\rule{2cm}{0mm}
}
\end{flushright}
 
\vspace{1.5cm}

\begin{center}

{\Large\bf \titletextCERN}



\vspace{1.5cm}

{\bf Dirk~Graudenz}$\,\!$\footnote[1]{\em Electronic
mail address: Dirk.Graudenz\char64{}cern.ch}$\!\!$\footnote[2]{\em
WWW URL: http://wwwcn.cern.ch/$\sim$graudenz/index.html}


{\it Theoretical Physics Division, CERN\\
1211 Geneva 23, Switzerland}

\end{center}

\vspace{1.0cm}

\begin{center}
{\bf Abstract}
\end{center}

\hspace{4mm}
\abstracttext

\vfill
\noindent
CERN--TH/\cernnr\\
\vdate

\clearpage
\setcounter{page}{1}

\renewcommand{\thefootnote}{\arabic{footnote}}
\setcounter{footnote}{0}

\fi
 

\ifnum\pltp=1

\begin{frontmatter}

\title{\titletextPL}

\author[CERN]{Dirk~Graudenz\thanksref{EMD1}\thanksref{EMD2}}

\address[CERN]{Theoretical Physics Division, CERN\\
1211 Geneva 23, Switzerland}

\thanks[EMD1]{\em Electronic
mail address: Dirk.Graudenz\char64{}cern.ch
}

\thanks[EMD2]{
\em WWW URL: http://wwwcn.cern.ch/$\sim$graudenz/index.html}

\begin{abstract}
\abstracttext
\end{abstract}

\end{frontmatter}

\fi

\section{Introduction}
The DESY H1 and ZEUS Collaborations have recently published results for
charged-particle production in deeply inelastic electron--proton scattering
at HERA \cite{1,2}.
In this paper, we calculate, in next-to-leading-order QCD perturbation theory,
the $x_F$-distribution in the current fragmentation region, 
employing the recent parametrization of charged-meson
fragmentation functions from Refs.~\cite{3,4}.
The comparison of theoretical and experimental results looks very encouraging
and suggests that a determination of the strong coupling constant~$\alpha_s$
via 
scaling violations of fragmentation functions might be feasible.
We give an estimate of the statistical error 
$\Delta\alpha_s^{\mbox{\scriptsize stat}}$ 
that can be expected for the HERA machine
runnning at design luminosity, and study the errors 
$\Delta\alpha_s^{\mbox{\scriptsize PDF}}$ and
$\Delta\alpha_s^{\mbox{\scriptsize scale}}$ 
due to the uncertainty coming from various
parton density parametrizations
and choices of the factorization scale, 
respectively. 
Except for the dependence on the parton density parametrization,
an $\alpha_s$-determination via scaling violations has the advantage
that, being based on a renormalization group equation, it is 
model-independent\dgfootnote{For example, 
in the case of (2+1)-jet production,
the experimental jet rates have to be ``corrected'' back to the parton level
by means of a fragmentation model, 
thus introducing an additional systematic uncertainty.}. A drawback is that
the size of the effect, being proportional to the logarithm of the 
energy scale, is small. 
Compared with $e^+e^-$\ annihilation,
HERA offers the unique possibility to perform the
required measurements at various energy scales at one machine without
changing the centre-of-mass energy.

The theoretical basis for the calculation of one-particle-inclusive cross 
sections is the 
factorization theorem of perturbative QCD (see, for example, 
Ref.~\cite{5} and references therein). 
For the process
$lP\rightarrow l^\prime h+X$
($l$, $l^\prime$ are the incoming and outgoing
leptons, $P$ is the incoming proton, $h$ is the observed hadron, and
$X$ denotes anything else in the hadronic final state\dgfootnote{
We identify the four-momenta of particles with their genuine names.}),
the cross section $\sigma$ 
can be written in the form 
\beqm{factform}
\sigma=\int \dd\xi\, f_{i/P}\left(\xi,\mu_f^2\right) 
\int \dd z\, D_{h/j}\left(z,\mu_D^2\right)
\,\sigma_{\mbox{\scriptsize hard}}^{ij}
\left(\xi,z,\mu_f^2,\mu_D^2,\mu_r^2\right),
\eeq
where $f_{i/P}$ and $D_{h/j}$ denote parton densities and
fragmentation functions, $\sigma_{\mbox{\scriptsize hard}}^{ij}$
is the mass-factorized hard scattering cross section,
$\xi$ and~$z$ are the momentum fractions of the incident 
parton~$i$ and of the observed hadron originating from the 
fragmenting parton~$j$, and~$\mu_f$, $\mu_D$~are the corresponding 
factorization scales. We have also indicated the renormalization 
scale~$\mu_r$. The phenomenological distribution functions~$f$ and~$D$
have to be taken from experiment. The hard scattering cross section 
$\sigma_{\mbox{\scriptsize hard}}$ has been calculated in Ref.~\cite{6}.
For the present numerical study, we have used an implementation of a 
recent recalculation \cite{7}.
In leading order, the cross section 
is given by the process of the naive parton model 
with a fragmentation function attached to the outgoing quark, see
Fig.~\ref{lodiagra}. 

\begin{figure}[htb] \unitlength 1mm
\begin{center}
\dgpicture{159}{39}

\put(46,0){\epsfigdg{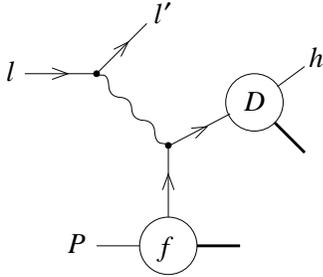}{width=42mm}} 

\end{picture}
\end{center}
\caption[]
{\labelmm{lodiagra} {\it Feynman diagram corresponding
to the leading-order contribution.
}}
\end{figure}

In next-to-leading order, virtual corrections have to be added
to the $\gamma^*q\rightarrow q$ diagram, 
and the QCD subprocesses $\gamma^*q\rightarrow qg$ and 
$\gamma^*g\rightarrow q\overline{q}$ have to be included.
After factorization of collinear singularities in the parton cross section 
and renormalization of 
the phenomenological distribution functions, the resulting 
one-particle-inclusive cross section is infrared-finite.
A comparison of theoretical and experimental 
$x_F$-distributions is done in the next section. 
The scales $\mu_f$, $\mu_D$ and $\mu_r$ are in principle arbitrary.
In the leading-order process under consideration, the only physical
scale related to the hard scattering process is the photon virtuality
$Q=\sqrt{-q^2}$, with $q$ the momentum of the exchanged photon. We will
therefore identify the factorization and renormalization scales with~$Q$, 
except for the case where we study the scale dependence explicitly.

We will consider only processes with an exchanged
virtual photon, and neglect the contributions from an exchanged $Z$~boson.
For the comparison with present-day experimental data, this is justified
by the restricted range in $Q$. For simplicity, the analysis
in Section~\ref{secthree}
of the error of a possible measurement of the strong coupling constant 
via scaling violations of fragmentation functions  
is done in this approximation
as well, even though $Q$ may reach up to $150\,\GeV$. We assume
this to be sufficient to achieve an estimate of the error, but it is clear
that for an experimental analysis at large~$Q$ 
the contributions for an exchanged
$Z$~boson have to be included. 

\section{Comparison with Experimental Data}
\label{sectwo}
In this section we compare the longitudinal momentum fraction
distributions\dgfootnote{We note that the present calculation does not permit
a next-to-leading-order comparison of transverse momentum spectra.}
\beqm{lspect}
\rho(x_F)=
\frac{1}{\sigma_{\mbox{\scriptsize tot}}}
\,\frac{\dd \sigma}{\dd x_F}
\eeq
recently published by 
the H1 \cite{1} and ZEUS \cite{2} Collaborations with the 
theoretical next-to-leading-order prediction.
Here $\sigma_{\mbox{\scriptsize tot}}$ is the total cross section
for the same cuts on the phase space of the outgoing lepton as for the
one-particle-inclusive cross section $\sigma$.
The variable $x_F$ is defined to be $2h_L/W$, where $h_L$ is the
component of the observed hadron's momentum along the direction of the 
exchanged virtual photon 
in the
hadronic centre-of-mass frame, 
and $W=\sqrt{(P+q)^2}$ is the total hadronic energy.
The current direction is defined by the condition $x_F\geq 0$.

\begin{figure}[htb] \unitlength 1mm
\begin{center}
\dgpicture{139}{103}

\put( 0,0){\epsfigdg{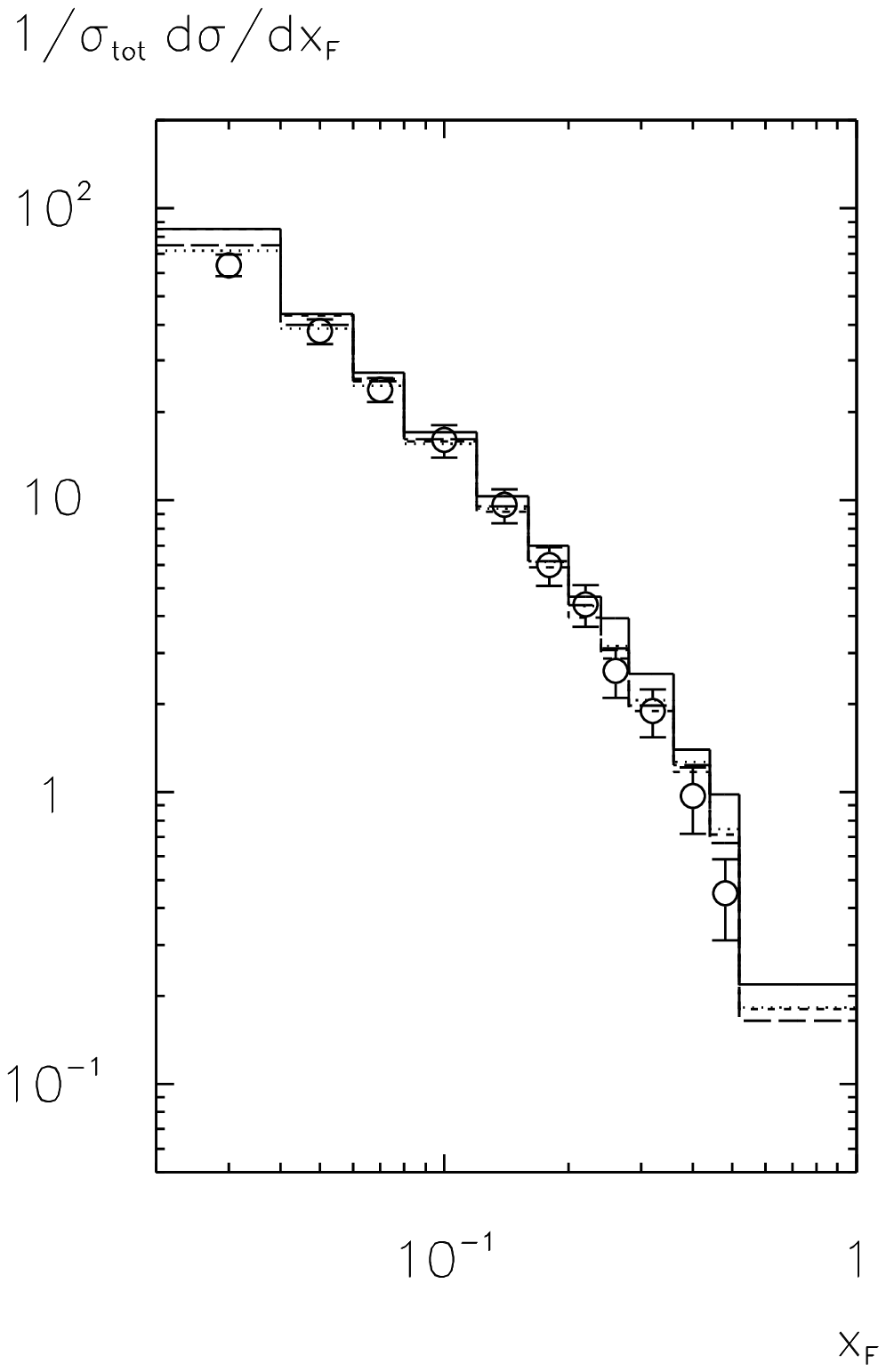}{width=66mm}} 
\put(70,0){\epsfigdg{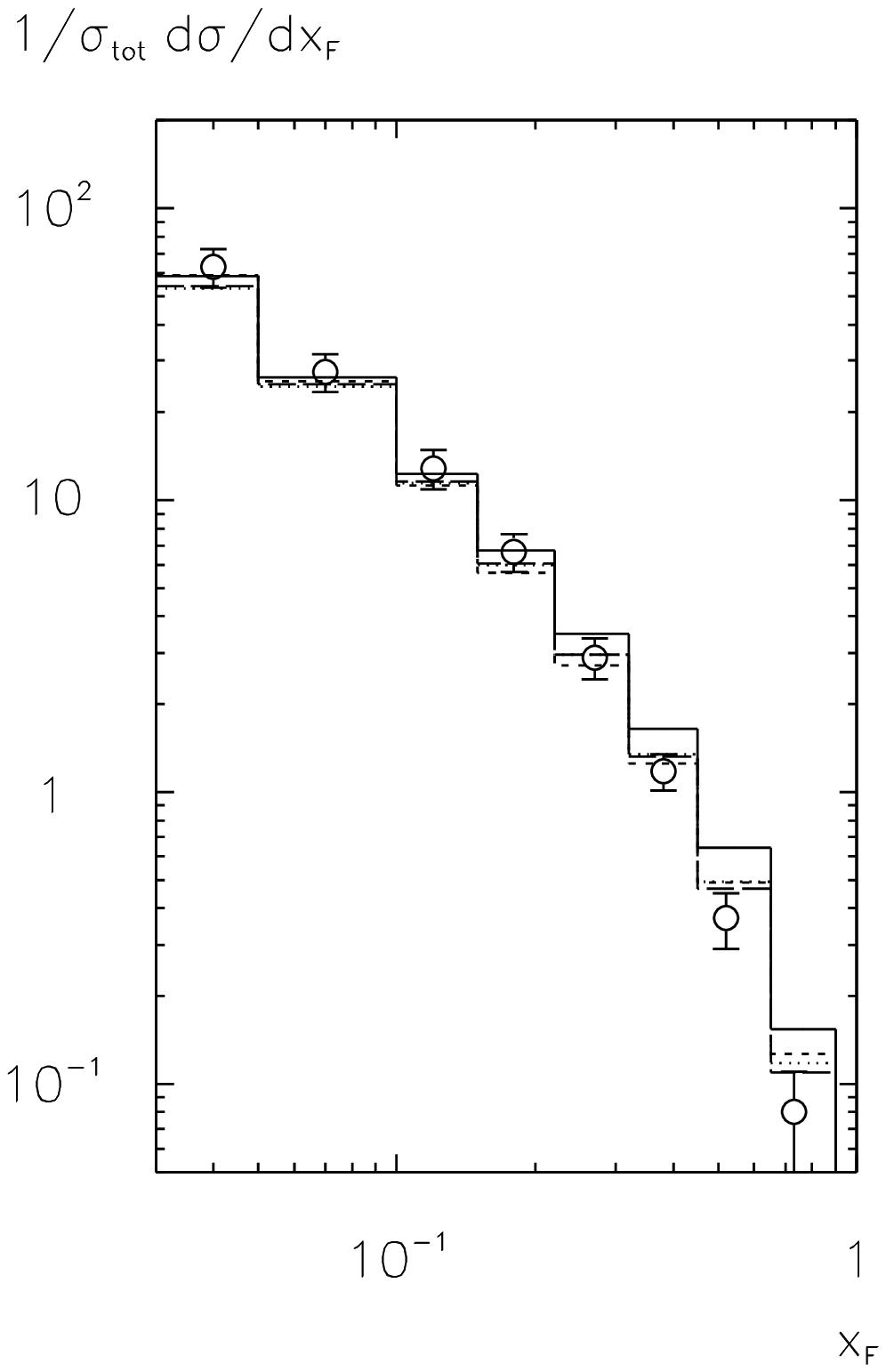}{width=66mm}} 

\put(10,0){\rm (a)}
\put(80,0){\rm (b)}

\end{picture}
\end{center}
\caption[]
{\labelmm{xfcmp} {\it 
Distributions in $x_F$ in comparison with H1 (a) and ZEUS (b) data.
The parton density parametrizations are:
GRV LO \mbox{\rm[\fullline]} (with leading-order matrix elements); GRV HO 
\mbox{\rm[\dashline]}, MRS A$^\prime$ \mbox{\rm[\dotline]},
CTEQ 3M \mbox{\rm[\longdashline]} (with next-to-leading-order matrix elements).
}}
\end{figure}

The HERA proton and lepton energies are $E_P=820\,\GeV$ and $E_l=26.7\,\GeV$,
respectively.
The cuts imposed in the experimental analyses are:
(a) 
H1 cuts: $10^{-4}\leq x_B\leq 10^{-2}$,
$Q\leq 10\,\GeV$, $54.77\,\GeV\leq W\leq 200\,\GeV$, 
$E_{l^\prime}\geq 14\,\GeV$,
$157^\circ\leq \vartheta_{l^\prime}\leq 172.5^\circ$;  
(b)
ZEUS cuts: $0.04\leq y\leq 0.85$,
$3.16\,\GeV\leq Q \leq 12.65\,\GeV$, 
$75\,\GeV\leq W\leq 175\,\GeV$, $E_{l^\prime}\geq 10\,\GeV$.
Here $x_B=Q^2/2Pq$, $y=Pq/Pl$, and $E_{l^\prime}$ and $\vartheta_{l^\prime}$
are the energy and polar 
angle of the outgoing lepton in the laboratory frame.
The experimental data are corrected from the limited detector acceptance
for the observed charged particle
to the full phase space.

Figure~\ref{xfcmp} shows the comparison of experimental data with the 
theoretical prediction for various sets of parton distribution
functions\dgfootnote{
The parton density parametrizations are:
GRV LO and GRV HO \cite{8}, MRS A$^\prime$ \cite{9} and
CTEQ 3M \cite{10}.
}. The employed value for $\Lambda_{\mbox{\scriptsize QCD}}$ is the
one from the parton distribution functions. Charm and bottom quarks are 
treated as massless flavours in the matrix element, and the flavour threshold
in the running coupling constant is assumed to be at the single heavy quark 
masses.
Except for very small and very large values of 
$x_F$, the agreement of theory and experiment is quite satisfactory.
The next-to-leading-order corrections are negative 
for most bins in $x_F$, and
bring the prediction in better agreement 
with the data,
in particular in the large-$x_F$ range.
Therefore, the QCD corrections are important to describe the data.
For large $x_F$, the theoretical prediction is, however, systematically
larger than the experimental result, but still
within about one standard deviation
of the data, depending on the parton density 
parametrization under consideration.
It is expected that the fragmentation function picture breaks down 
at small $x_F$, because soft fragmentation effects, such as particle
production between the current and remnant jets, become important. 
Moreover, the fragmentation function fit in 
Refs.~\cite{3,4} only takes into account data for 
$0.1\leq x_F\leq 0.8$. Outside of
these bounds, the fragmentation functions are
extrapolated. We also wish to note that the experimental data include
all produced charged hadrons, whereas the theoretical prediction, due to a 
lack of a suitable set of fragmentation functions, is restricted to 
charged pions and kaons.

\begin{figure}[htb] \unitlength 1mm
\begin{center}
\dgpicture{139}{103}

\put( 0,0){\epsfigdg{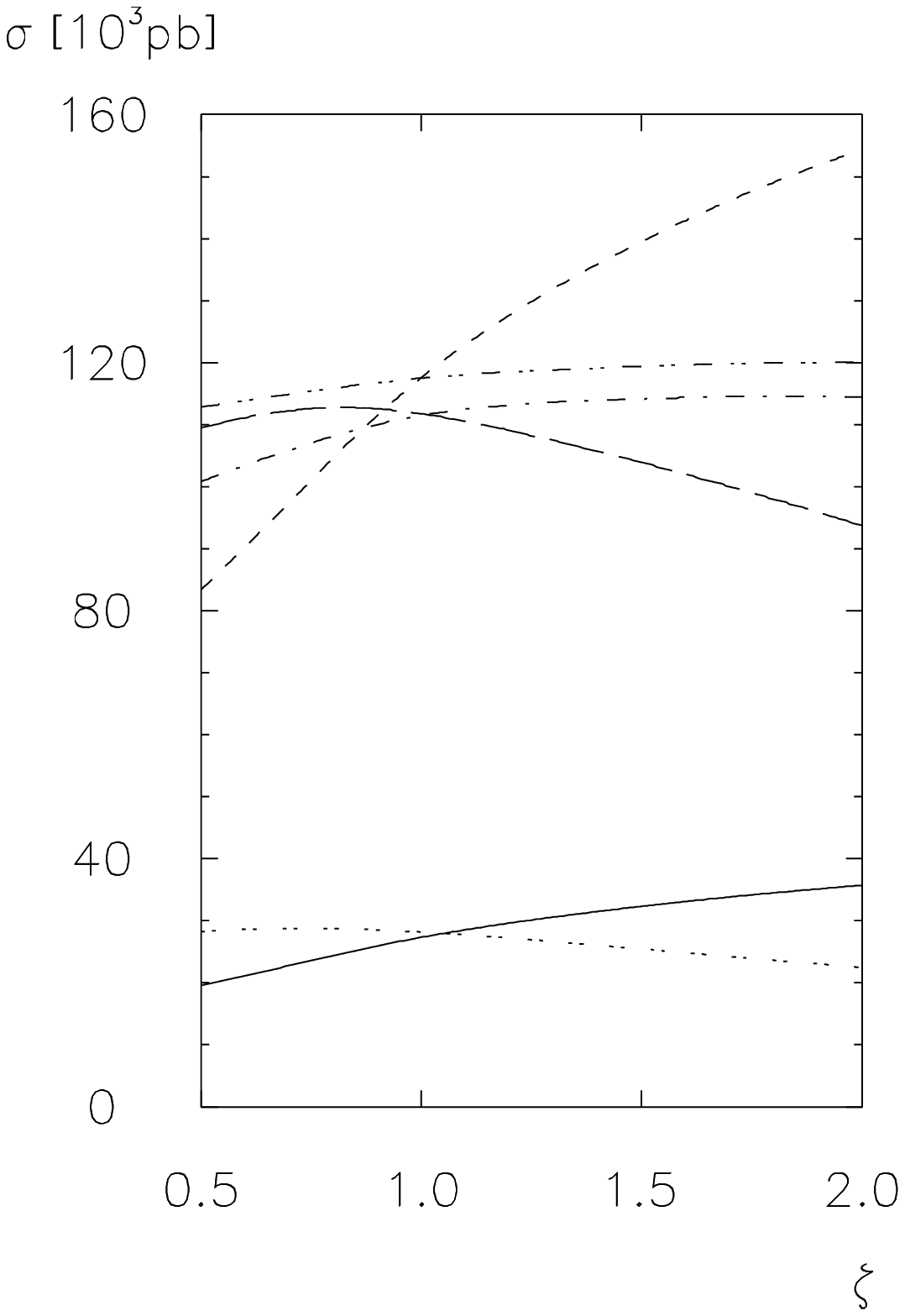}{width=66mm}} 
\put(70,0){\epsfigdg{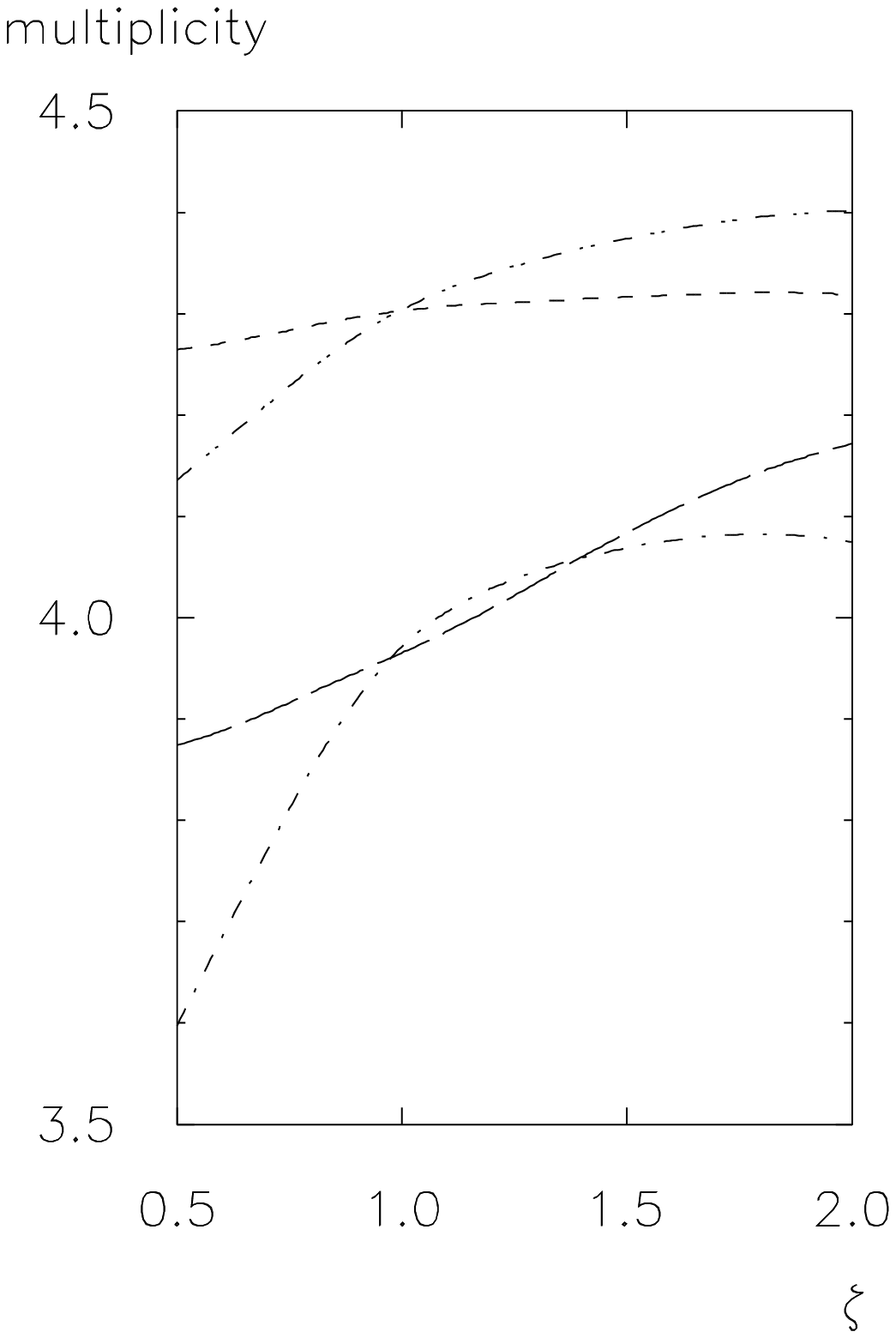}{width=64mm}} 

\put(10,0){\rm (a)}
\put(80,0){\rm (b)}

\end{picture}
\end{center}
\caption[]
{\labelmm{scalefig} {\it 
Scale dependence of cross sections (a) and multiplicity (b) for the ZEUS cuts.
The particular scale set to $\zeta Q$ is given by: (a)
$\mu_f$ \mbox{(\dashline)}, $\mu_D$ \mbox{(\dashdotdotline)}
in leading order and
$\mu_f$ \mbox{(\longdashline)},
$\mu_D$ \mbox{(\dashdotline)}
in next-to-leading order for the one-particle-inclusive cross section $\sigma$,
$\mu_f$ in leading \mbox{(\fullline)} and next-to-leading \mbox{(\dotdotline)}
order for the total cross section $\sigma_{\mbox{\scriptsize tot}}$; 
(b) $\mu_f$ \mbox{(\dashline)}, $\mu_D$ \mbox{(\dashdotdotline)}
in leading order and               
$\mu_f$ \mbox{(\longdashline)},                        
$\mu_D$ \mbox{(\dashdotline)}                              
in next-to-leading order for the multiplicity.
The other scales are fixed to be equal to~$Q$.
}}
\end{figure}

We finish this section with some remarks concerning the scale dependence
of the theoretical prediction.
Figure~\ref{scalefig} shows the dependence of 
the cross sections and of the charged multiplicity
$\sigma/\sigma_{\mbox{\scriptsize tot}}$, for the cuts of the ZEUS analysis,
as a function of the factorization
scales $\mu_f$ and $\mu_D$, varying individually 
in the form $\zeta Q$,
the other scales being kept fixed at $Q$. 
A minimum value of $2\,\GeV$ for the factorization scales is required.
The parton densities are given by the GRV parametrizations.
For the one-particle-inclusive cross section,
the dependence on $\mu_f$ is 
much flatter in next-to-leading order than in 
leading order, whereas the dependence on $\mu_D$ is 
slightly larger in next-to-leading
order (although it develops an extremum). 
Including only the scale-compensating terms from the next-to-leading-order 
matrix element proportional to a product
of a splitting function and a logarithm in $\mu_D^2$, 
the $\mu_D$-dependence in next-to-leading order is much smaller than the
one in leading order. The enhanced $\mu_D$-dependence is therefore a genuine
higher-order effect.
The $\mu_f$-dependence of the total cross section
is reduced in next-to-leading order as well.
Although the scale dependence of the absolute cross sections looks
quite reasonable, the scale dependence of the multiplicity
is larger in next-to-leading order than in leading order 
(Fig.~\ref{scalefig}b). A similar study shows that the overall scale dependence
of the multiplicity
is significantly reduced for larger values of $Q$.
The impact of the scale uncertainty on an $\alpha_s$-measurement will
be studied in the next section.
The dependence on the renormalization scale, not shown in the figures,
is very small, because the running coupling constant enters the matrix elements
only in next-to-leading order.

\section{The Strong Coupling Constant via Scaling Violations?}
\label{secthree}
The good agreement of the $x_F$-distributions of theory and data, as discussed
in the previous section, is an encouraging sign for the feasibility
of a measurement of the strong coupling constant via scaling violations
of fragmentation functions.
The scale evolution of fragmentation functions is governed by a
renormalization group equation \cite{11}\dgfootnote{
The evolution kernels $K_{k\leftarrow j}\left(u,\alpha_s\right)$ 
in next-to-leading order can be found in Ref.~\cite{12}; see
also Ref.~\cite{13}.
}:
\beqm{Devol}
\frac{\partial D_{h/j}(z,\mu_D^2)}{\partial \ln \mu_D^2}=
\frac{\alpha_s(\mu_D^2)}{2\pi}\int_z^1\frac{\dd u}{u}\,
K_{k\leftarrow j}\left(u,\alpha_s(\mu_D^2)\right)
\,D_{h/k}\left(\frac{z}{u},\mu_D^2\right).
\eeq
A multiparameter fit of fragmentation functions for partons into 
charged hadrons thus permits a determination of the fundamental
QCD parameter $\Lambda_{\mbox{\scriptsize QCD}}$ (or equivalently the
strong coupling constant $\alpha_s\left(M_Z^2\right)$ at the mass of 
the $Z$~boson), if the measurement is performed for several scales $\mu_D$.
Since the scale evolution depends only on the logarithm of the
scale, it is necessary to include a wide range of scales in the fit. 
The photon virtuality $Q$ is the only scale at hand in the process
under consideration\dgfootnote{The total hadronic energy~$W$ is not directly
related to the hard scattering process, because it contains contributions
from the proton remnant jet.
}, 
and consequently the analysis will require high 
statistics, because of the rapidly falling cross section with increasing $Q$.

We study here three potential sources of errors: the statistical error due
to a limited number of events, in particular at large~$Q$; the error 
coming from the spread of the theoretical cross sections due to
various parametrizations of parton densities, and the uncertainty
due to the choice of the factorization scale $\mu_D$.
The error estimate presented here can, of course, 
not replace an analysis based on a $\chi^2$-fit; it is rather intended
to give a first assessment of the sizes of the various uncertainties.

The procedure to obtain an estimate of the statistical error of $\alpha_s$ is 
the following. We fix the
fragmentation functions at a scale of $\mu_0=2\,\GeV$ as the leading-order
parametrization of Refs.~\cite{3,4}\dgfootnote{
We have to make an assumption of this kind in order to get an estimate
of the dependence of the $x_F$-distribution on the value of the
strong coupling constant. Since later on we take ratios of the distributions
at two different scales (we are only interested in the slope depending
on $Q$), it is not really important at which scale 
we identify the distributions. 
}. 
We then evolve this input with two 
different values 
for $\Lambda_{\mbox{\scriptsize QCD}}^{(4)}$
of \{a\} $0.1\,\GeV$ and \{b\} $0.2\,\GeV$.
The corresponding 
$x_F$-distributions
$\rho^{\{a\}}$ and $\rho^{\{b\}}$ 
are determined for these two sets of fragmentation functions. With the
cuts of the ZEUS data analysis unmodified, except for the cut on~$Q$, 
we determine the distributions
$\rho^{\{a1\}}$, $\rho^{\{a2\}}$, $\rho^{\{a3\}}$,
$\rho^{\{b1\}}$, $\rho^{\{b2\}}$, $\rho^{\{b3\}}$
for three bins\dgfootnote{It turns out that the dependence on the
parton densities is unacceptably large for smaller values of $Q$
due to the large spread of parametrizations at small~$x$. Bin~\{1\} is the
$Q$-range of the original ZEUS analysis. The other two bins are chosen 
such that one of them includes data only at very large $Q$. 
We expect that bin optimization will lead to a smaller overall error
by balancing 
statistical and systematic errors; this is, however, beyond the scope
of the present paper.} in $Q$: 
\{1\} $[3.16,12.6]\,\GeV$, 
\{2\} $[12.6,100]\,\GeV$ and
\{3\} $[100,150]\,\GeV$. 
The ratios $\lambda^{\{21\}}=\rho^{\{2\}}/\rho^{\{1\}}$ and 
$\lambda^{\{32\}}=\rho^{\{3\}}/\rho^{\{2\}}$ for an arbitrary coupling constant 
$\alpha_s$ (taken at the mass of the $Z$~boson)
are assumed to depend linearly 
on $\alpha_s$: 
\beqm{ratio}
\lambda^{\{ij\}}=\lambda^{\{aij\}}
+\frac{\lambda^{\{bij\}}-\lambda^{\{aij\}}}
      {\alpha_s^{\{b\}}-\alpha_s^{\{a\}}}\,
\left(\alpha_s-\alpha_s^{\{a\}}\right).
\eeq
For a given luminosity, and under the assumption of a Gaussian error for the
event numbers, this allows us to estimate the statistical error $\epsilon_k$
of $\alpha_s$
for every bin $k$ in $x_F$. These individual errors are then combined 
into a total error~$\epsilon$
according to 
$\epsilon=1/\sqrt{\sum_k(1/\epsilon_k^2)}$.
To obtain explicit numerical values, we use the CTEQ 3L
parametrization \cite{10} for the
parton densities 
(for simplicity, we work in leading order).
The integrated luminosity is assumed to be $250\,\mbox{pb}^{-1}$
(the HERA design luminosity, per experiment, integrated over five years).
For the analysis based on bins~\{1\} and~\{2\}, we obtain a statistical error 
of $\alpha_s(M_Z^2)$ of
$\pm 0.0007$, and for the bins~\{2\} and~\{3\}, the statistical error is
$\pm 0.027$. 

Another problem of the extraction of $\alpha_s$ is the dependence of 
the theoretical
cross sections on the parton density parametrization. Taking the ratio of 
one-particle-inclusive and total cross sections leads to a substantial
cancellation, but a certain residual dependence remains. 
To estimate the size of this effect, we determine the spread of the results
for $\alpha_s(M_Z^2)$ 
depending on the next-to-leading-order parton densities 
from Refs.~\cite{8,9,10}. For the bins~\{1\} and~\{2\}, 
the spread is
$\pm 0.017$, and for the bins~\{2\} and~\{3\}, the spread is
$\pm 0.005$.
Future global fits of parton densities including improved HERA data
should reduce this systematic uncertainty.

Finally, we discuss the dependence of $\alpha_s$ on the choice of the 
factorization scale~$\mu_D$. To obtain an estimate, the ratios~$\lambda$ 
are determined for the three choices $Q/2$, $Q$ and $2Q$ of this
scale. The change of cross section has for consequence a variation in 
the extracted $\alpha_s(M_Z^2)$-value of $\pm 0.013$ and 
$\pm 0.011$ for the
combinations of the bins \{1\},~\{2\} and \{2\},~\{3\}, respectively.
\section{Summary and Conclusions}

We have calculated, in next-to-leading order, the 
$x_F$-distribution for the production of charged mesons in deeply inelastic
scattering at HERA. The comparison with experimental data is satisfactory.
The dependence of the theoretical cross section prediction on the 
factorization scale $\mu_f$
is considerably reduced, whereas the scale dependence
of the charged multiplicity is increased. 
The $\mu_D$-dependence of the next-to-leading-order cross sections, 
although slightly larger
than in leading order, is small. The cross section results can therefore
be considered to be 
reliable for a quantitative test of QCD. 

As a particular example, we have 
estimated the statistical and systematic errors (due to theoretical 
uncertainties) for an extraction of the strong coupling constant via scaling 
violations of fragmentation functions. We find, for the HERA machine
running at design 
luminosity for five years, 
$\Delta\alpha_s^{\mbox{\scriptsize stat}}\approx$ 
$\pm 0.0007$/$\pm 0.027$,
$\Delta\alpha_s^{\mbox{\scriptsize PDF}}\approx$
$\pm 0.017$/$\pm 0.005$ 
and 
$\Delta\alpha_s^{\mbox{\scriptsize scale}}\approx$ 
$\pm 0.013$/$\pm 0.011$ 
for two specific sets of cuts.
It is expected that there is still some room for improvement by means
of bin optimization.
We have not considered experimental systematic uncertainties.
They will also contribute to the error on $\alpha_s$.
The dominant systematic error in the experimental $x_F$-distributions
comes from the boost of the particle momenta from the laboratory system 
into the hadronic centre-of-mass
system. For a measurement of $\alpha_s$ it may therefore be advantageous
to define a suitable observable in the laboratory system, to circumvent this
source of systematic uncertainty.

Compared with the present error $\Delta\alpha_s=0.006$ of the 
world average, the errors estimated here are large.
It should be kept in mind, however, that a reduction of 
$\Delta\alpha_s^{\mbox{\scriptsize PDF}}$ can be expected
because of improved fits of parton densities, in particular in the
small-$x$ region. 
It might also be possible to reduce the dependence on the parton density
by performing the analysis in the Breit frame \cite{14}.
A reduction of 
$\Delta\alpha_s^{\mbox{\scriptsize scale}}$ would
require a higher-order calculation or a physical scheme for the choice
of the factorization scales.
A measurement
of $\alpha_s$ at HERA via scaling violations of fragmentation functions
is worth doing 
because it is an independent quantitative test of QCD and, more important, 
because it 
complements other methods for an $\alpha_s$-determination in 
deeply inelastic scattering, such 
as the measurement via (2+1)-jet rates and via scaling violations of structure 
functions.

\medskip
\medskip
\medskip

\noindent
{\bf Acknowledgements}

\medskip
\medskip

\noindent
I wish to thank N.~Brook, M.~Kuhlen and N.~Pavel for discussions and
for comments on the manuscript, Ch.~Berger and T.~Doyle
for discussions, 
and J.~Binnewies for clarifying remarks concerning
the parametrizations of Refs.~\cite{3,4}.
This work was supported in part by a Habilitandenstipendium of the
Deutsche Forschungsgemeinschaft.

\newcommand{\bibbeginlong}{\begin{thebibliography}{XXX\,1988\,\,}}
\newcommand{\bibbeginshort}{\begin{thebibliography}{XX}}

\newcommand{\bibitemb}[4]{\begin{sloppy}
{\bibitem[#1]{#1} {#2}, {#4.}}
\end{sloppy}}

\newcommand{\nbibitemb}[4]{}

\bibbeginshort

\bibitemb{1}
      {I.~Abt et al.\ (H1 Collaboration)}
      {}
      {Z.~Phys.\ \underline{C63} (1994) 377}

\bibitemb{2}
      {M.~Derrick et al.\ (ZEUS Collaboration)}
      {}
      {Z.~Phys.\ \underline{C70} (1996) 1}

\bibitemb{3}
      {J.~Binnewies, B.A.~Kniehl and G.~Kramer}
      {Next-to-Leading Order Fragmentation Functions
       for Pions and Kaons}
      {Z.~Phys.\ \underline{C65} (1995) 471}

\bibitemb{4}
      {J.~Binnewies, B.A.~Kniehl and G.~Kramer}
      {}
      {Phys.\ Rev.\ \underline{D52} (1995) 4947}

\bibitemb{5} 
      {J.C.~Collins, D.E.~Soper and G.~Sterman}
      {Factorization of Hard Processes in QCD}
      {in: {\it Perturbative Quantum Chromodynamics,} ed.\ A.H.~Mueller,
       World Scientific (Singapore, 1989)}

\bibitemb{6}
      {G.~Altarelli, R.K.~Ellis, G.~Martinelli and S.Y.~Pi}
      {Processes Involving Fragmentation Functions Beyond the 
       Leading Order in QCD}
      {Nucl.\ Phys.\ \underline{B160} (1979) 301}

\bibitemb{7}
       {D.~Graudenz}
       {}
       {{\it Heavy-Quark Production in the Target Fragmentation Region,} 
        CERN--TH/96--52, in preparation} 

\bibitemb{8}
       {M.~Gl\"{u}ck, E.~Reya and A.~Vogt}
       {Dynamical Parton Distributions of the Proton
        and Small~$x$ Physics}
       {Z.~Phys.\ \underline{C67} (1995) 433} 

\bibitemb{9}
      {A.~Martin, R.~Roberts and J.~Stirling}
      {}
      {Phys.\ Rev.\ \underline{D51} (1995) 4756}

\bibitemb{10}
      {CTEQ Collaboration}
      {Global QCD Analysis and the CTEQ Parton Distributions}
      {Phys.\ Rev.\ \underline{D51} (1995) 4763}

\bibitemb{11} 
      {R.~Baier and K.~Fey}
      {Finite Corrections to Quark Fragmentation Functions in  
       Perturbative QCD}
      {Z.~Phys.\ \underline{C2} (1979) 339}

\bibitemb{12}
      {G.~Curci, W.~Furmanski and R.~Petronzio}
      {Evolution of Parton Densities Beyond Leading Order}
      {Nucl.\ Phys.\ \underline{B175} (1980) 27}

\bibitemb{13}
      {P.~Nason and B.R.~Webber}
      {}
      {Nucl.\ Phys.\ \underline{B421} (1994) 473}

\bibitemb{14} 
      {N.~Brook and T.~Doyle}
      {}
      {private communication}

\end{thebibliography}

\end{document}